\documentclass[11pt]{article}

\usepackage{a4wide}
\usepackage{color}

\usepackage[bookmarks=true,bookmarksnumbered=true,setpagesize=false]{hyperref}
\usepackage{enumerate}
\usepackage{amsmath,amssymb,amsthm}

\usepackage{textcomp,mathcomp} 

\usepackage{graphicx}
\usepackage{braket}
\usepackage{url}

\usepackage[OMLmathsfit]{isomath} 
\newcommand{\mt}{\mathsfit} 
\usepackage{mathrsfs}

\newcommand{\h}{\hat}
\newcommand{\tx}{\text}

\usepackage{fontawesome5} 
\newcommand{\fs}{\text{\faStarO}}

\usepackage{comment}

\renewcommand{\t}{\text{t}}

\let\origoverrightarrow\overrightarrow

\usepackage{mathabx}

\let\overrightarrow\origoverrightarrow

\newcommand{\lip}[1]{\left\lgroup#1\right\rgroup}

\newcommand{\bp}{\begin{pmatrix}}
\newcommand{\ep}{\end{pmatrix}}
\newcommand{\bb}{\begin{bmatrix}}
\newcommand{\eb}{\end{bmatrix}}
\newcommand{\bmat}[1]{\begin{bmatrix}#1\end{bmatrix}}

\DeclareMathOperator{\diag}{diag}

\DeclareMathOperator{\PLC}{PLC}

\newcommand{\df}{\text{d}}

\newcommand{\bs}{\boldsymbol}

\newcommand{\ub}{\underbrace}

\newcommand{\bh}[1]{\boldsymbol{\hat{#1}}}

\newcommand{\al}[1]{\begin{align}#1\end{align}}
\newcommand{\als}[1]{\begin{align*}#1\end{align*}}

\newcommand{\ab}[1]{\left|#1\right|}

\newcommand{\paren}[1]{\left(#1\right)}
\newcommand{\pn}[1]{\left(#1\right)}
\newcommand{\sqbr}[1]{\left[#1\right]}
\newcommand{\br}[1]{\left\{#1\right\}} 

\newcommand{\Ab}[1]{\bigl|#1\bigr|}

\newcommand{\Pn}[1]{\bigl(#1\bigr)}

\newcommand{\autospace}{%
  \mathchoice%
    {\!}
    {\!}
    {}
    {}
}
\newcommand{\fn}[1]{\autospace\paren{#1}} 
\newcommand{\Fn}[1]{\autospace\Pn{#1}} 

\newcommand{\wt}{\widetilde}
\newcommand{\nn}{\nonumber\\}

\newcommand{\p}{\partial}

\newcommand{\red}[1]{{\color[cmyk]{0,0.8,1,0}#1}}
\newcommand{\green}[1]{{\color[cmyk]{0.97,0,0.75,0}#1}}
\newcommand{\blue}[1]{{\color[cmyk]{1,0.5,0,0}#1}}

\newcommand{\orange}[1]{{\color[cmyk]{0,0.5,1,0}#1}}
\newcommand{\black}[1]{{\color[cmyk]{0,0,0,1}#1}}

\usepackage{fancybox}

\usepackage{amsthm}
\theoremstyle{definition}

\newcommand{\ov}{\over}

\newcommand{\mc}{\mathcal}

\newcommand{\lv}[1]{\overrightarrow{#1}}

\newcommand{\nab}{\bs\nabla}

\newcommand{\sP}{\s_\text{P}}
\newcommand{\xP}{x_\text{P}}
\renewcommand{\P}{\text{P}}

\newcommand{\pr}{\prime}

\newcommand{\sn}{\blue{\s_n}}

\newcommand{\bx}{\orange{\bs x}}
\newcommand{\lx}{\orange{\lv x}}
\newcommand{\xz}{\orange{x^0}}
\newcommand{\oxi}{\orange{x^i}}
\newcommand{\oxj}{\orange{x^j}}

\newcommand{\snsx}{\s_n^\star(\lx)}

\newcommand{\restrict}[2]{\left.#1\right|_{#2}}

\newcommand{\s}{s}

\begin{document}
\title{
Covariant Electromagnetism in Past-Light-Cone Formalism
}
\author{
Daiju Nakayama,\thanks{E-mail: \tt 42.daiju@gmail.com}{}\ {}
Kin-ya Oda,\thanks{E-mail: \tt odakin@lab.twcu.ac.jp}{}\ {}
and Koichiro Yasuda \thanks{E-mail: \tt yasuda@physics.ucla.edu}\bigskip\\
\it\normalsize$^*$ e-Seikatsu Co., Ltd., 5-2-32, Tokyo 106-0047, Japan
\\
\it\normalsize$^\dagger$ Department of Information and Mathematical Sciences,\\
\it\normalsize Tokyo Woman's Christian University, Tokyo 167-8585, Japan\\
\it\normalsize$^\ddag$ Department of Physics and Astronomy, University of California, Los Angeles,\\
\it\normalsize California 90095-1547, USA\\
}
\maketitle
\begin{abstract}\noindent
We present a manifestly covariant formulation of relativistic electromagnetism, focusing on the computation of electromagnetic fields from moving charges in a manifestly Lorentz-covariant manner. The electromagnetic field at a given spacetime point is determined by the motion of point charges at the intersection of their worldlines with the past light cone of the spacetime point, ensuring causal consistency. This formalism provides a manifestly covariant generalization of the Li\'enard-Wiechert potentials and allows direct implementation on computers. We compare our formulation with standard textbook approaches and analyze its behavior in various physical limits.
\end{abstract}

\newpage

\tableofcontents

\newpage

\section{Introduction}

Electromagnetism, being a fundamental interaction in nature, is inherently relativistic. The formulation of Maxwell’s equations in a Lorentz-covariant form played a crucial role in the development of special relativity~\cite{Einstein:1905ve}. While the standard covariant formulation expresses the electromagnetic field through the field-strength tensor, standard methods for computing fields from moving charges often rely on expressions that are explicitly evaluated in a preferred reference frame. Such approaches obscure the direct relationship between charge dynamics and the fields at a given spacetime event, making it difficult to analyze their transformation properties in a fully covariant manner.

A well-known example of such frame-dependent formulations is the derivation of the electromagnetic field from a moving point charge using the Li\'enard-Wiechert potential; see, e.g., Refs.~\cite{Landau:1975pou,Sunakawa}. This potential provides a standard solution to Maxwell's equations for a given charge trajectory, but its conventional derivation involves an implicit choice of an observer's frame and does not manifestly preserve Lorentz covariance. As a result, computing the field observed in a different inertial frame typically requires additional Lorentz transformations, rather than being directly obtained from a fully covariant formulation.

To achieve a fully Lorentz-covariant description of electromagnetic fields from moving charges, we adopt a formulation based on past-light-cone (PLC) structure; see Ref.~\cite{10.1093/ptep/ptx127} for an implementation of the PLC formalism on computers. In this approach, the electromagnetic field at a given spacetime point is determined by the motion of point charges at the intersection of their worldlines with the past light cone of the spacetime point. Unlike conventional derivations, which often involve evaluating the field in a specific frame before transforming it to another, our approach maintains manifest covariance throughout by constructing the field-strength tensor directly from source dynamics in a Lorentz-invariant manner.

Our formulation provides a manifestly covariant generalization of the Li\'enard-Wiechert potential, directly incorporating the causal structure of electrodynamics. By expressing the electromagnetic field in terms of the PLC structure, we avoid the need to specify an observer's frame in intermediate steps, maintaining Lorentz covariance throughout. This contrasts with conventional derivations, where the field is often computed in a preferred frame and subsequently transformed. Our approach provides a direct and systematic method to compute the field-strength tensor from charge worldlines, offering a framework suitable for both theoretical analysis and numerical implementation.

This paper is organized as follows. In Sec.~\ref{elemag review section}, we formulate the equations governing the motion of charged particles and describe the covariant framework for field generation. In Sec.~\ref{covariant formalism section}, we derive the field-strength tensor directly from charge worldlines, demonstrating the manifestly covariant nature of our approach. In Sec.~\ref{Conclusion}, we summarize our results and discuss potential applications. In Appendix~\ref{general solution section}, we review the standard derivation of the Green's function solution to Maxwell's equations in the Lorenz gauge. In Appendix~\ref{sec:Useful formulae}, we summarize basic derivative formulae used throughout the paper. In Appendix~\ref{sec:Spacetime differentiation on past light cone and of retarded vector potential}, we verify the consistency of spacetime differentiation on the past light cone and present explicit expressions for the spacetime derivatives of the vector potential.

\section{Relativistic electromagnetism}\label{elemag review section}

\begin{figure}[tp]
\centering
\includegraphics[width=0.4\textwidth]{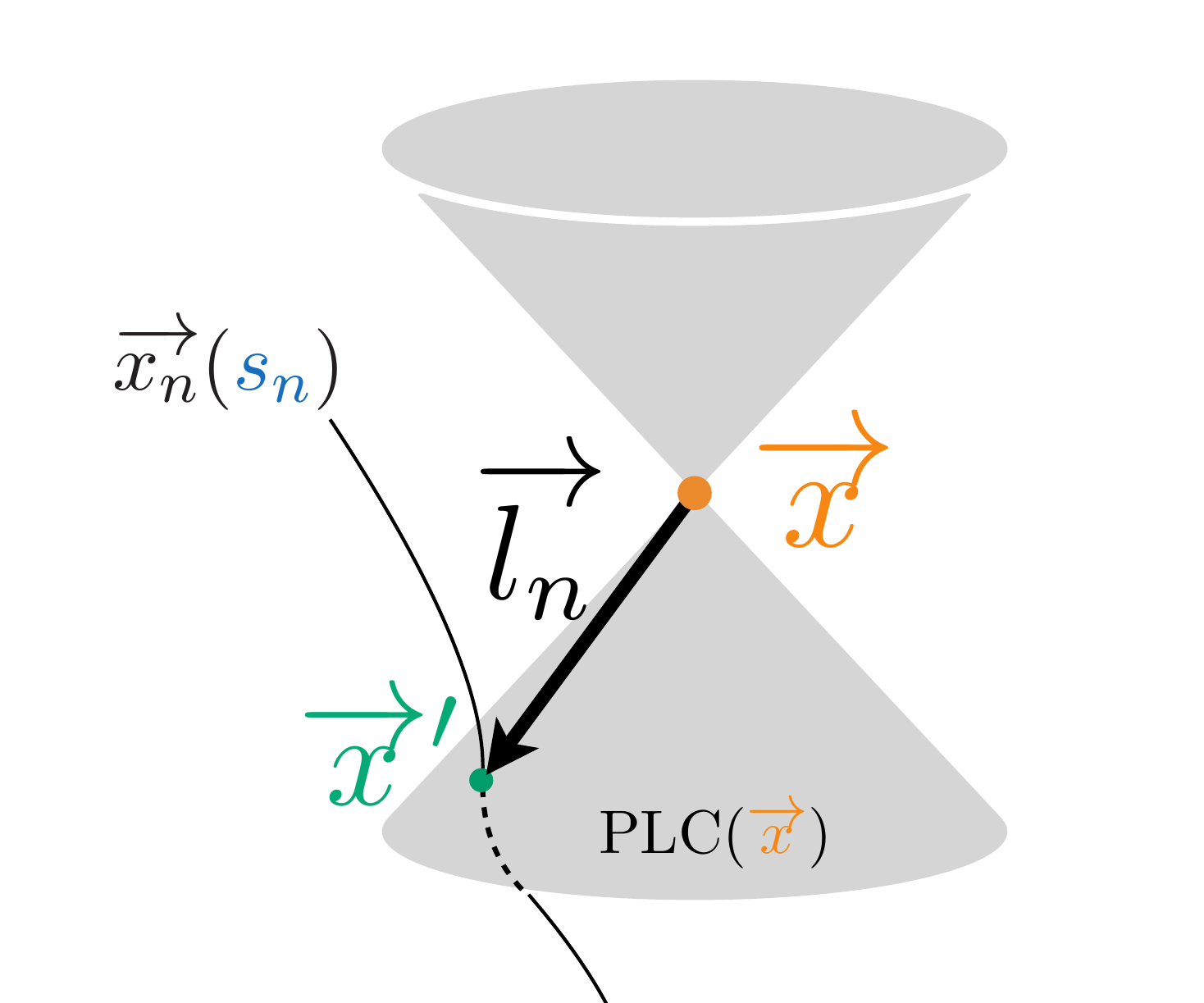}
\caption{Schematic illustration:
$\protect\lx$ is the spacetime point at which the electromagnetic field is evaluated;
$\PLC\fn{\protect\lx}$ is its past light cone;
$\protect\lv{x_n}\fn{\sn}$ parametrizes the worldline of the $n$th charge;
$\green{\protect\lv x'}$ is a spacetime location of a charge that influences the field at the observation point;
and $\protect\lv{l_n}$ is the chargeward vector.
\label{intersection figure}
}
\end{figure}

This section briefly reviews the standard framework of relativistic electromagnetism to establish notation and conventions for the covariant formulation developed in the following sections.

Spacetime coordinates throughout this paper are written as
\al{
\lv x
	&=	\pn{x^0,\bs x}
	=	\pn{x^0,x^1,x^2,x^3},
}
where $x^0=ct$, with $c$ being the speed of light.
We denote spacetime vectors (with upper indices) using large arrows, and spatial vectors in boldface.

\subsection{Point charges}
The (dimensionless) Lorentz-covariant velocity of the $n$th charge is defined by
\al{
\lv{u_n}
	&:=	{\df\lv{x_n}\ov\df\blue{s_n}},&
u_n^\mu
	&=	{\df x_n^\mu\ov\df\blue{s_n}},
}
where Greek indices $\mu,\nu,\dots$ run over $0,1,2,3$, and $\blue{s_n}$, referred to as the ``proper time distance'' of the $n$th charge, is defined by\footnote{
This paper does not employ the natural unit $c=1$, so that length and time retain distinct dimensions, which motivates the term ``proper time distance''. The non-natural unit facilitates practical computer implementations discussed elsewhere.
}
\al{
\df\blue{s_n}=\sqrt{-\eta_{\mu\nu}\df x_n^\mu\df x_n^\nu},
	\label{proper time distance defined}
}
under the metric convention
\al{
\bmat{\eta_{\mu\nu}}_{\mu,\nu=0,1,2,3}=\diag\fn{-1,1,1,1}.
	\label{metric convention}
}
For clarity, some quantities are highlighted in color throughout this paper.
The covariant acceleration of the $n$th charge is defined as
\al{
\lv{\alpha_n}
	&:=	{\df\lv{u_n}\ov{\df\sn}}
	=	{\df^2\lv{x_n}\ov{\df\sn}^2},&
\alpha_n^\mu
	&=	{\df u_n^\mu\ov\df\sn}
	=	{\df^2 x_n^\mu\ov{\df\sn}^2}.
	\label{covariant acceleration}
}
Note that the acceleration carries the dimension of inverse length here.

From definition~\eqref{proper time distance defined}, we find
\al{
\lip{\lv{u_n}}^2
	&=	-1,&
u_n^0
	&=	\sqrt{1+\bs u_n^2},
	\label{normalization of u}
}
where $\lip{\lv{u_n}}^2 := \lv{u_n}\cdot\lv{u_n}$, with $\lv{u_n}\cdot\lv{u_n} := u_n^\mu\eta_{\mu\nu}u_n^\nu$.
Thus, $\lv{u_n}$ has only three independent components $\bs u_n=\pn{u_n^1,u_n^2,u_n^3}$, while $u_n^0$ is determined by them.
Taking the derivative of the first equation in Eq.~\eqref{normalization of u}, we obtain
\al{
\lv{\alpha_n}\cdot\lv{u_n}
	&=	0,&
\alpha_n^0
	&=	{\bs\alpha_n\cdot\bs u_n\ov\sqrt{1+\bs u_n^2}}.
	\label{temporal acceleration}
}
This shows that the temporal component of the covariant acceleration, $\alpha_n^0$, is not an independent degree of freedom.

Given the motion of the charges, we can express the charge density $\rho\fn{\green{\lv x}}$ and the 3D current density $\bs j\fn{\green{\lv x}}$ at a spacetime point $\green{\lv x}$ as
\al{
\rho\fn{\green{\lv x}}
	&=	\sum_nq_n\int\df\sn \,\delta^4\Fn{\green{\lv x}-\lv{x_n}\fn{\sn }},
		\label{rho given}\\
\bs j\fn{\green{\lv x}}
	&=	\sum_nq_n\int\df\sn \,\delta^4\Fn{\green{\lv x}-\lv{x_n}\fn{\sn }} c\bs u_n\fn{\sn }.
		\label{j given}
}
Here, $\bs u_n\fn{\sn}$ is dimensionless, hence the factor $c$ above.

Given the electromagnetic field strength $F$, defined below in Eq.~\eqref{field strength defined}, the equation of motion under the Lorentz force takes the manifestly Lorentz-covariant form:
\al{
m_n c{\df u_n^\mu\fn{\sn}\ov\df\sn }
	&=	q_nF^{\mu\nu}\Fn{\lv{x_n}\fn{\sn}}u_\nu\fn{\sn}.
}
Using matrix notation, introduced in Eqs.~\eqref{matrix F} and \eqref{transformation of F} below, this equation becomes
\al{
m_n c{\df\lv{u_n}\fn{\sn}\ov\df\sn }
	&=	q_n\mt F\Fn{\lv{x_n}\fn{\sn}}\,\mt\eta\,\lv u\fn{\sn}.
		\label{eom simple form}
}
This form is more convenient for numerical implementation.

\subsection{Relativistic Maxwell's equations}

The charge and current densities~\eqref{rho given} and \eqref{j given}, respectively, determine the electromagnetic fields via Maxwell's equations at each spacetime point $\lx$:
\al{
\nab\cdot\bs E\fn{\lx}
	&=	{\rho\fn{\lx}\ov\epsilon_0},&
\nab\times\bs B\fn{\lx}
	&=	{\bs j\fn{\lx}\ov\epsilon_0c^2}
		+{1\ov c^2}c\p_0\bs E\fn{\lx},
		\label{matter interaction}\\
\nab\cdot\bs B\fn{\lx}
	&=	0,&
\nab\times\bs E\fn{\lx}
	&=	-c\p_0\bs B\fn{\lx}.\label{purely electromagnetic}
}
where
\al{
\p_\mu
	&:=	{\p\ov\p x^\mu},&
c\p_0
	&=	c{\p\ov\p x^0}={\p\ov\p t},
}
and $\epsilon_0$ is the electric constant (vacuum permittivity).

The homogeneous Maxwell equations~\eqref{purely electromagnetic} are automatically solved by the scalar and 3D vector potentials $\phi\fn{\lx}$ and $\bs A\fn{\lx}$:
\al{
\bs B\fn{\lx}
	&=	\nab\times\bs A\fn{\lx},&
\bs E\fn{\lx}
	&=	-\nab\phi\fn{\lx}-c\p_0\bs A\fn{\lx},
		\label{BE in terms of Aphi}
}
and we define the (Lorentz-covariant) vector potential $\lv A$ and current density $\lv j$,
\al{
\lv A\fn{\lx}
	&=	\Pn{A^\mu\fn{\lx}}_{\mu=0,\dots,3}
	:=	\pn{{\phi\fn{\lx}\ov c},\bs A\fn{\lx}},\\
\lv j\fn{\lx}
	&=	\Pn{j^\mu\fn{\lx}}_{\mu=0,\dots,3}
	:=	\Pn{c\rho\fn{\lx},\bs j\fn{\lx}},
}
as well as the field strength tensor
\al{
F_{\mu\nu}\fn{\lx}
	&:=	\p_\mu A_\nu\fn{\lx}-\p_\nu A_\mu\fn{\lx}.
		\label{field strength defined}
}
The electromagnetic fields can be written in terms of the field-strength tensor: Eq.~\eqref{BE in terms of Aphi} reads
\al{
E_i\fn{\lx}
	&=	cF_{i0}\fn{\lx}
	=	cF^{0i}\fn{\lx},\nn
B_i\fn{\lx}
	&=	{1\ov2}\epsilon_{ijk}F_{jk}\fn{\lx}
	=	{1\ov2}\epsilon_{ijk}F^{jk}\fn{\lx},
		\label{electromagnetic field}
}
where $\epsilon_{ijk}$ is the totally antisymmetric tensor with $\epsilon_{123}=1$ and the Einstein summation convention is used for both spatial indices $i,j,\dots$ running over $1,2,3$ and the spacetime indices $\mu,\nu,\dots$. Note that $i,j,\dots$ do not require the distinction between upper and lower indices due to the metric convention~\eqref{metric convention}.

Maxwell's equations~\eqref{matter interaction} can be recast in covariant form as
\al{
\p_\nu F^{\mu\nu}\fn{\lx}
	&=	{j^\mu\fn{\lx}\ov\epsilon_0c^2}.
			\label{matter equation of motion}
}
In the Lorenz gauge, Maxwell's equation~\eqref{matter equation of motion} reduces to
\al{
-\Box A^\mu\fn{\lx}
	&=	{j^\mu\fn{\lx}\ov\epsilon_0c^2};
	\label{Maxwell's equation in Lorenz gauge}
}
see Appendix~\ref{sec:Rewriting Maxwell's equation in Lorenz gauge} for its derivation.

We may write down the general solution to Eq.~\eqref{Maxwell's equation in Lorenz gauge}, under appropriate boundary conditions, as
\al{
A^\mu\fn{\lx}
	&=	{1\ov 4\pi\epsilon_0c^2}\int_{\PLC\fn{\lx}} \df^3\green{\bs{x'}} {j^\mu\fn{\green{\lv x'}} \ov \left|\bx - \green{\bs x'}\right|},
	\label{general solution of A}
}
or, more explicitly,
\al{
	A^\mu\fn{\lx}
	&= {1\ov 4\pi \epsilon_0 c^2}\int\df^4\green{\lv x'} 
		\delta\fn{\xz-\green{x^{\pr0}} -  \ab{\bx - \green{\bs{x^\pr}}}}
	{j^\mu\fn{\green{\lv x'}}\ov\ab{\bx - \green{\bs{x^\pr}}}}.
		\label{A concrete}
}
A derivation of these expressions is reviewed in Appendix~\ref{sec:Green's function method for solving vector potential}.

Physically, $\green{\lv x'}$ corresponds to a spacetime point in the past at which a charge contributes to the electromagnetic field observed at $\lx$; see Fig.~\ref{intersection figure}.  
The delta function restricts $\green{\lv x'}$ to lie on the past light cone $\PLC\fn{\lx}$ by enforcing the condition $\xz - \green{x^{\pr0}} = \ab{\bx - \green{\bs x'}} \geq 0$.

\section{Covariant formalism for field strength from point charges}\label{covariant formalism section}
This section is the main body of this paper.
We compute how the relativistic motion of a charged particle affects the electromagnetic field in the future. In other words, we compute a fully relativistic expression of the Li\'enard-Wiechert potential in terms of covariant quantities only.
Our main goal is to find out the field strength at $\orange\lx$ in terms of the positions and velocities of point charges on $\PLC\fn{\orange\lx}$.

\subsection{Worldlines and covariant current density}

To compute the electromagnetic field at a spacetime point $\lx$, we consider the contributions from all charged particles whose worldlines intersect $\PLC\fn{\lx}$. Each contribution is determined solely by the position and velocity of the particle at the point of intersection; see Fig.~\ref{intersection figure}.

Given the worldlines $\lv{x_n}\fn{\blue{s_n}}$ of the charged particles, the covariant current density at a spacetime point $\green{\lv x'}$ is given by
\al{
\lv j\fn{\green{\lv x'}} &= \sum_ncq_n\int\df\sn\,\delta^4\fn{\green{\lv x'}-\lv{x_n}\fn{\sn }}\lv{u_n}\fn{\sn}, \label{current j}
}
where $q_n$ is the charge of the $n$th particle. Here, the factor $c$ is included to give the current density the dimension of $\pn{\tx{charge}}/\pn{\tx{length}}^2\pn{\tx{time}}$.

\subsection{Master equation for vector potential}

We now derive an explicit expression for the vector potential $A^\mu\fn{\lx}$ in terms of the positions and velocities of point charges on $\PLC\fn{\lx}$; the final result is given in Eq.~\eqref{final true master equation for vector potential}.

Substituting Eq.~\eqref{current j} into the general solution~\eqref{general solution of A}, or more explicitly Eq.~\eqref{A concrete}, we obtain
\al{
A^\mu\fn{\orange{\lx}}
	&=	\sum_n{q_n\ov4\pi\epsilon_0c}
		\int\df\sn \,u_n^\mu\fn{\sn }
		\int \df^4\green{\lv x'} {\delta\fn{\xz-\green{x^{\pr0}}-\ab{\bx-\green{\bs x'}}} \ov \ab{\bx -\green{\bs x'}}}
		\delta^4\fn{\green{\lv x'}-\lv{x_n}\fn{\sn }}\nn
	&=	\sum_n{q_n\ov4\pi\epsilon_0c}
		\int\df\sn \,u_n^\mu\fn{\sn }
		{\delta\fn{\xz-x_n^0\fn{\sn}-\ab{\bx-\bs x_n\fn{\sn}}}
			\ov
			\ab{\bx-\bs x_n\fn{\sn}}}.
		\label{A start explicit}
}

To evaluate this integral, we first define the \emph{chargeward vector}, which connects the field point $\lx$ to the position of the $n$th charge:
\al{
\lv{l_n}\fn{\sn,\lx}
	&:=	\lv{x_n}\fn{\sn}-\lx,&
\bmat{l_n^0\fn{\sn,\xz}\\ \bs l_n\fn{\sn,\bx}}
	&:=	\bmat{x_n^0\fn{\sn}-\xz \\ \bs x_n\fn{\sn }-\bx}.
	\label{ln given}
}
We also introduce the corresponding unit spatial vector:
\al{
\bh l_n\fn{\sn,\bx}
	=	{\bs l_n\fn{\sn,\bx}\ov\ab{\bs l_n\fn{\sn,\bx}}}.
}
Using the chargeward vector, we define a modified gamma factor, which involves the derivative of the argument of the delta function in Eq.~\eqref{A start explicit} with respect to the proper time distance:
\al{
\gamma_n\fn{\sn,\bx}
	&:=	{\df x_n^0\fn{\sn}\ov\df\sn}+{\p\ab{\bs l_n\fn{\sn,\bx}}\ov\p\sn}.
		\label{gamma_n firstly defined}
}
With the derivative formula~\eqref{d abs l d tau} in Appendix~\ref{sec:Useful formulae}, it becomes
\al{
\gamma_n\fn{\sn,\bx}
	=	u_n^0\fn{\sn }+\bh l_n\fn{\sn,\bx}\cdot\bs u_n\fn{\sn }.
	\label{gamma_n defined}
}
The positivity $\gamma_n\fn{\sn,\bx} \geq 0$ follows from $u_n^0 = \sqrt{1+\bs u_n^2} \geq \ab{\bs u_n}$.

Employing the chargeward vector and the modified gamma factor introduced above, we can now perform the integration in Eq.~\eqref{A start explicit}, arriving at
\al{
A^\mu\fn{\lx}
	&=	\sum_n{q_n\ov4\pi\epsilon_0c}
		{u_n^\mu\Fn{\snsx}\ov\gamma_n\Fn{\snsx,\bx}\ab{\bs l_n\Fn{\snsx,\bx}}},
		\label{final true master equation for vector potential}
}
where $\snsx$ is the solution to
\al{
\xz-x_n^0\fn{\sn}
	&=	\ab{\bx-\bs x_n\fn{\sn}}
		\label{yet another trivial equation}
}
with respect to $\sn$.

Physically, $\snsx$ corresponds to the proper time distance at which the worldline of the $n$th charge intersects the past light cone of the field point $\lx$. The expression~\eqref{final true master equation for vector potential} provides a causal formulation of the vector potential written in terms of Lorentz covariant quantities, making it suitable for practical computations.

\subsection{Field strength}\label{time derivative section}
We obtain the spacetime derivatives of the vector potential:
\al{
\p_0A^\mu\fn{\lx}
	&=	\sum_n{q_n\ov4\pi\epsilon_0c}
		\sqbr{
			{\alpha_n^\mu\ov\gamma_n^2\ab{\bs l_n}}
			-{u_n^\mu\ov\gamma_n^3\ab{\bs l_n}}\pn{
				\alpha_n^0
				+\bh l_n\cdot\bs\alpha_n
				+{\bs u_n\cdot\pn{\bs u_n+u_n^0\bh l_n}\ov\ab{\bs l_n}}}
			},\\
\p_iA^\mu\fn{\lx}
	&=	\sum_n{q_n\ov4\pi\epsilon_0c}
		\sqbr{
			{\h l_n^i\alpha_n^\mu\ov\gamma_n^2\ab{\bs l_n}}
			+{u_n^iu_n^\mu\ov\gamma_n^2\ab{\bs l_n}^2}
			-{\h l_n^iu_n^\mu\ov\gamma_n^3\ab{\bs l_n}}\pn{\alpha_n^0+\bh l_n\cdot\bs\alpha_n-{1\ov\ab{\bs l_n}}}
			};
}
see Appendix~\ref{sec:Spacetime differentiation on past light cone and of retarded vector potential} for detailed derivation.
Here and hereafter, we omit the arguments for the quantities $\lv{u_n}\fn{\sn}$, $\lv{\alpha_n}\fn{\sn}$, $\lv{l_n}\fn{\sn,\lx}$, and $\gamma_n\fn{\sn,\lx}$, as well as 
the restriction to the past light cone $\PLC\fn{\lx}$, i.e., the condition $\sn=\snsx$ is assumed unless otherwise stated.
Everything is now written in terms of covariant quantities.

The components of field strength, $F_{i0}=-F_{0i}=F^{0i}=-F^{i0}$ and $F_{ij}=-F_{ji}=-F^{ji}=F^{ij}$, are now
\al{
F^{0i}\fn{\lx}
	&=	-\p_0A^i\fn{\lx}-\p_iA^0\fn{\lx}\nn
	&=	\sum_n{q_n\ov4\pi\epsilon_0c\ab{\bs l_n}}
		\sqbr{
			\h l^i{
						u_n^0
							\pn{
								\bh l_n\cdot\bs\alpha_n
								-{1\ov\ab{\bs l_n}}
								}
						-\alpha_n^0\pn{\bh l_n\cdot\bs u_n}
					\ov\gamma_n^3
						}
			+u_n^i{
			\alpha_n^0+\bh l_n\cdot\bs\alpha_n
			-{1\ov\ab{\bs l_n}}
			\ov\gamma_n^3
			}
			-{\alpha_n^i\ov\gamma_n^2}
		},\label{F0i}\\
F^{ij}\fn{\lx}
	&=	\p_iA^j\fn{\lx}-\p_jA^i\fn{\lx}\nn
	&=	\sum_n{q_n\ov4\pi\epsilon_0c\ab{\bs l_n}}\sqbr{
			{\h l_n^i\alpha_n^j-\h l_n^j\alpha_n^i\ov\gamma_n^2}
			-{\h l_n^iu_n^j-\h l_n^ju_n^i\ov\gamma_n^3}\pn{\alpha_n^0+\bh l_n\cdot\bs\alpha_n-{1\ov\ab{\bs l_n}}}
			},
			\label{Fij}
}
where Eq.~\eqref{gamma_n defined} is used in the last step of Eq.~\eqref{F0i}.
This expression for the field strength is one of our main results. It can be directly used to compute the electric and magnetic fields and their Lorentz transformation in a concrete implementation, as discussed below.

Given the field strength, the electric and magnetic fields can be derived from Eq.~\eqref{electromagnetic field}.
For an actual implementation in a computer program, one may write the field strength as an anti-symmetric matrix $\mt F$ whose $\mu,\nu$ components are given by the upper-indexed counterparts:
\al{
\bmat{\mt F\fn{\lx}}^{\mu,\nu}
	&=	F^{\mu\nu}\fn{\lx}.
	\label{matrix F}
}
Then its Lorentz transformation law under the coordinate transformation $\lx\to\orange{\lx'}=\mt\Lambda\lx$ is
\al{
\mt F
	&\to	\mt F'=\mt\Lambda\mt F\mt\Lambda^\t.
	\label{transformation of F}
}
Accordingly, the Lorentz transformation for the electromagnetic fields are
\al{
E^i
	&\to	E^{\pr i}
	=	\red cF^{\pr0i}
	=	\red c\bmat{\mt\Lambda\mt F\mt\Lambda^\t}^{0,i},\\
B^i
	&\to	B^{\pr i}
	=	{1\ov2}\epsilon_{ijk}F^{\pr jk}
	=	{1\ov2}\epsilon_{ijk}\bmat{\mt\Lambda\mt F\mt\Lambda^\t}^{j,k}.
}

\subsection{Electromagnetic field}

The result above can be directly implemented in computer programs to derive the Lorentz transformation law for electromagnetic fields. Therefore, it is not necessary to present explicit formulae for the electromagnetic fields themselves, which are, after all, not Lorentz covariant. Nonetheless, we proceed to derive expressions written solely in terms of covariant quantities, in order to facilitate comparison with the existing literature.

The electromagnetic fields~\eqref{electromagnetic field} are now
\al{
E^i\fn{\lx}
	&=	cF^{0i}\fn{\lx}\nn
	&=	\sum_n{q_n\ov4\pi\epsilon_0\ab{\bs l_n}}
		\sqbr{
			\h l_n^i{
						u_n^0
							\pn{
								\bh l_n\cdot\bs\alpha_n
								-{1\ov\ab{\bs l_n}}
								}
						-\alpha_n^0\pn{\bh l_n\cdot\bs u_n}
					\ov\gamma_n^3
						}
			+u_n^i{
			\alpha_n^0+\bh l_n\cdot\bs\alpha_n
			-{1\ov\ab{\bs l_n}}
			\ov\gamma_n^3
			}
			-{\alpha_n^i\ov\gamma_n^2}
		},
			\label{electric field obtained so far}\\
B^i\fn{\lx}
    &=  {1\ov 2}\sum_{j,k}\epsilon^{ijk}F^{jk}\fn{\lx}\nn
	&=	\sum_n{q_n\ov4\pi\epsilon_0c\ab{\bs l_n}}
			\br{\bh l_n\times\sqbr{
				{\bs\alpha_n\ov\gamma_n^2}
				-{\bs u_n\ov\gamma_n^3}\pn{
                        \alpha_n^0
                        +\bh l_n\cdot\bs\alpha_n
                        -{1\ov\ab{\bs l_n}}}
				}
			}_i;\label{Our B field}
}
see Eq.~\eqref{ln given} for the definition of the chargeward vector $\lv{l_n}$.\footnote{
As usual, we used $\epsilon_{ijk}\h l_n^j\alpha_n^k-\h l_n^k\alpha_n^j=\Pn{\bh l\times\bs \alpha_n}_i$, etc.
}
%
From the above expression, we immediately see
\al{
c\bs B_n\fn{\lx}
	&=	-\bh l_n\fn{\lx}\times\bs E_n\fn{\lx}.
		\label{BE relation}
}
where $\bs B_n\fn{\lx}$, $\bs E_n\fn{\lx}$ are the fields created by the n-th charge, and we have abbreviated as $\bh l_n\fn{\lx}:=\bh l_n\Fn{\snsx,\bx}$.
Accordingly, the Poynting vector reads
\al{
\bs S
	&:=	\epsilon_0c^2\bs E\fn{\lx}\times\bs B\fn{\lx}\nn
    &=  \epsilon_0c^2\sum_{n,n'}\bs E_n\fn{\lx}\times\bs B_{n'}\fn{\lx},
		\label{Poynting vector}
}
which represents the energy flux (power flow) of the electromagnetic field.\footnote{
Retaining only the $n = n'$ term allows us to isolate the influence of the $n$th particle:
\als{
\bs S_n
	&=	-\epsilon_0c\sqbr{\bs E_n^2\fn{\lx}\bh l_n\fn{\lx}-\pn{\bs E_n\fn{\lx}\cdot\bh l_n\fn{\lx}}\bs E_n\fn{\lx}},
}
In the full expression \eqref{Poynting vector}, however, cross terms involving different particles also appear.
}

\subsection{Comparison with literature}
We now rewrite our result in terms of non-covariant quantities and show that it reproduces well-known textbook expressions.
For this purpose, it is convenient to define the non-covariant acceleration for each point charge $q_n$:\footnote{
Following Eq.~\eqref{covariant acceleration}, non-covariant accelerations are likewise expressed in units of $\pn{\tx{length}}^{-1}$.
}
\al{
\bs w_n
	&:=	{\df^2\bs x_n\ov\pn{\df x_n^0}^2}
	=	{1\ov c}{\df\bs v_n\ov\df x_n^0},
}
which can be written in terms of the covariant quantities as
\al{
\bs w_n
	&=	{\bs\alpha_n-\pn{\bs\alpha_n\cdot{\bs u_n\ov u_n^0}}{\bs u_n\ov u_n^0}\ov\pn{u_n^0}^2}
	\ \pn{=	{\bs\alpha_n-\alpha_n^0{\bs u_n\ov u_n^0}\ov\pn{u_n^0}^2}},
}
where we used Eqs.~\eqref{normalization of u} and \eqref{temporal acceleration}.

After some computation, we obtain
\al{
\bs E\fn{\lx}
	&=	\sum_n{q_n\ov4\pi\epsilon_0\ab{\bs l_n}\pn{1+{\bh l_n\cdot\bs u_n\ov u_n^0}}^3}\sqbr{
			\pn{\bh l_n\cdot\bs w_n}\pn{\bh l_n+{\bs u_n\ov u_n^0}}
			-\pn{1+{\bh l_n\cdot\bs u_n\ov u_n^0}}\bs w_n
			-{\bh l_n+{\bs u_n\ov u_n^0}\ov\ab{\bs l_n}\pn{u_n^0}^2}
			}.
			\label{electric field final}
}
If one instead uses the \emph{fieldward vector},
\al{
\bh n_n\fn{\lx}:=-\bh l_n\fn{\lx},
	\label{fieldward vector}
}
from the point charge to the reference point~$\lx$, one may rewrite this as
\al{
\bs E\fn{\lx}
	&=	\sum_n{q_n\ov4\pi\epsilon_0\ab{\bs l_n}\pn{1-{\bh n_n\cdot\bs u_n\ov u_n^0}}^3}\sqbr{
			\pn{\bh n_n\cdot\bs w_n}\pn{\bh n_n-{\bs u_n\ov u_n^0}}
			-\pn{1-{\bh n_n\cdot\bs u_n\ov u_n^0}}\bs w_n
			+{\bh n_n-{\bs u_n\ov u_n^0}\ov\ab{\bs l_n}\pn{u_n^0}^2}
			}.
			\label{electric field rewritten}
}
Hereafter, we mainly use the fieldward vector $\bh n_n$ to indicate direction, and the chargeward vector $\ab{\bs l_n}$ to denote the distance between the field and charge.

We have explicitly checked that our result~\eqref{electric field rewritten} coincides with the non-covariant expression in Eq.~(3.29) in Chapter~9 in Ref.~\cite{Sunakawa}:
\al{
\bs E\fn{\lx}
	&=	\sum_n{q_n\ov4\pi\epsilon_0}\pn{
			{\pn{\bh n_n-{\bs v_n\ov c}}\pn{1-{\bs v_n^2\ov c^2}}\ov\pn{1-\bh n_n\cdot{\bs v_n\ov c}}^3\ab{\bs l_n}^2}
			+{\bh n_n\times\pn{\pn{\bh n_n-{\bs v_n\ov c}}\times{1\ov c}{\df\bs v_n\ov\df x_n^0}}\ov\pn{1-\bh n_n\cdot{\bs v_n\ov c}}^3\ab{\bs l_n}}
			},
}
where the (dimensionful) non-covariant velocity is defined as usual:
\al{
\bs v_n
	&:=	c{\df\bs x_n\ov\df x_n^0},
}
satisfying $1-{\bs v_n^2\ov c^2}=1/\pn{u_n^0}^2$.
Otherwise, in the large $\ab{\bs l_n}$ limit, the last term in the square brackets of our result~\eqref{electric field rewritten} drops out and, using $\bs v_n= c\bs u_n/u_n^0$, we reproduce the approximate Eq.~(73.8) in Ref.~\cite{Landau:1975pou}:\footnote{
Recall the vector calculus formula:
\als{
\pn{\bs A\times\pn{\bs B\times\bs C}}_i
	&=	\epsilon_{ijk}A_j\pn{\sum_{l,m}\epsilon_{klm}B_lC_m}
	=	\pn{\sum_k\epsilon_{ijk}\epsilon_{lmk}}A_jB_lC_m\nn
	&=	\pn{\delta_{il}\delta_{jm}-\delta_{im}\delta_{jl}}A_jB_lC_m
	=	\Pn{\bs B\pn{\bs A\cdot\bs C}-\pn{\bs A\cdot\bs B}\bs C}_i,
}
where the last step is valid only when $\bs A$ and $\bs B$ are commutative as in the current consideration.
}
\al{
\bs E\fn{\lx}
	&=	\sum_n{q_n\ov4\pi\epsilon_0\ab{\bs l_n}}{
		\bh n_n\times\sqbr{\pn{\bh n_n-{\bs v_n\ov c}}\times\bs w_n}
		\ov\pn{1-\bh n_n\cdot{\bs v_n\ov c}}^3}
		+\mc O\fn{\ab{\bs l_n}^{-2}}.
}

\subsection{Various limits}
Let us examine the behavior of the fields in various limiting cases.
First, the non-relativistic limit $\bs u_n\to0$ yields $u_n^0\to1$ and $\bs w_n\to\bs\alpha_{n\perp}$, where $\bs\alpha_{n\perp}:=\bs\alpha_n-\bh n_n\pn{\bh n_n\cdot\bs\alpha_n}$ is the component of acceleration perpendicular to the line of sight.
In this limit, the contribution from the charge~$q_n$ to the electromagnetic fields, $\bs E_n$ and $\bs B_n$, becomes
\al{
\bs E_n\fn{\lx}
	&\to	{q_n\ov4\pi\epsilon_0\ab{\bs l_n}}\pn{
			-\bs\alpha_{n\perp}
			+{\bh n_n\ov\ab{\bs l_n}}
			},\\
\bs B_n\fn{\lx}
	&\to	{q_n\ov4\pi\epsilon_0c\ab{\bs l_n}}\pn{
			-\bh n_n\times\bs\alpha_{n\perp}
			}.
}
In the further limit $\bs\alpha_{n\perp}\to0$, we recover the ordinary Coulomb's law: $\bs E_n\fn{\lx}\to{q_n\ov4\pi\epsilon_0}{\bh n_n\ov\ab{\bs l_n}^2}$ and $\bs B_n\fn{\lx}\to0$.

Second, in the absence of acceleration, $\bs\alpha_n\to0$, so that $\alpha_n^0\to0$ and $\bs w_n\to0$, the contribution from $q_n$ becomes
\al{
\bs E_n\fn{\lx}
	&\to	{q_n\ov4\pi\epsilon_0\ab{\bs l_n}^2\pn{u_n^0-\bh n_n\cdot\bs u_n}^3}
		\pn{u_n^0\bh n_n-\bs u_n},\\
\bs B_n\fn{\lx}
	&\to	{q_n\ov4\pi\epsilon_0c\ab{\bs l_n}^2\pn{u_n^0-\bh n_n\cdot\bs u_n}^3}
		\pn{-\bh n_n\times\bs u_n}.
}
We see that further taking the non-relativistic limit $\bs u_n\to0$ again recovers Coulomb's law.
It is also noteworthy that the leading $\ab{\bs l_n}^{-1}$ term vanishes in the absence of acceleration.

Third, in the ultra-relativistic limit $\ab{\bs u_n}\gg1$, the electromagnetic fields produced by the charge $q_n$ take the form
\al{
\bs E_n\fn{\lx}
	&\to	{q_n\ov4\pi\epsilon_0\ab{\bs l_n}\pn{1-\bh n_n\cdot\bh u_n}^3\ab{\bs u_n}^2}
		\sqbr{
			\pn{\bh n_n\cdot\bs\alpha_{n\wt\perp}+{1\ov\ab{\bs l_n}}}\pn{\bh n_n-\bh u_n}
			-\pn{1-\bh n_n\cdot\bh u_n}\bs\alpha_{n\wt\perp}
			},
			\label{E ultra-relativistic}\\
\bs B_n\fn{\lx}
	&\to	-{q_n\ov4\pi\epsilon_0c\ab{\bs l_n}\pn{1-\bh n_n\cdot\bh u_n}^3\ab{\bs u_n}^2}
		\br{\bh n_n\times\sqbr{
			\pn{\bh n_n\cdot\bs\alpha_{n\wt\perp}+{1\ov\ab{\bs l_n}}}
				\bh u_n
			+\pn{1-\bh n_n\cdot\bh u_n}\bs\alpha_{n\wt\perp}
			}},
			\label{B ultra-relativistic}
}
where $\bs\alpha_{n\wt\perp}:=\bs\alpha_n-\pn{\bs\alpha_n\cdot\bh u_n}\bh u_n$ is the component of acceleration perpendicular to the velocity. In the further limit $\bh u_n\to\bh n_n$, where the motion is directed along the line of sight, both the numerator and denominator tend to zero, and the sub-leading ultra-relativistic terms omitted in Eqs.~\eqref{E ultra-relativistic} and \eqref{B ultra-relativistic} contribute to the limiting expressions~\eqref{E limit} and \eqref{B limit} below.

Finally, in the limit where the observation direction $\bh n_n$ approaches the velocity direction $\bh u_n$, the fields due to the charge $q_n$ take the form
\al{
\bs E_n\fn{\lx}
	&\to	{q_n\ov4\pi\epsilon_0\ab{\bs l_n}\pn{1-{\ab{\bs u_n}\ov u_n^0}}^2}\sqbr{
			-\bs w_{n\perp}
			+{\bh n_n\ov\ab{\bs l_n}\pn{u_n^0}^2}
			},\\
\bs B_n\fn{\lx}
	&\to	-{q_n\ov4\pi\epsilon_0c\ab{\bs l_n}\pn{1-{\ab{\bs u_n}\ov u_n^0}}^2}\bh n_n\times\bs w_{n\perp}.
}
If we further take the ultra-relativistic limit $\ab{\bs u_n}\to\infty$, we find
\al{
\bs E_n\fn{\lx}
	&\to	{4\ab{\bs u_n}^2q_n\ov4\pi\epsilon_0\ab{\bs l_n}}\pn{
			-\bs\alpha_{n\perp}
			+{\bh n_n\ov\ab{\bs l_n}}
			},
			\label{E limit}\\
\bs B_n\fn{\lx}
	&\to	-{4\ab{\bs u_n}^2q_n\ov4\pi\epsilon_0c\ab{\bs l_n}}\,
			\bh n_n\times\bs\alpha_{n\perp},
			\label{B limit}
}
where $\bs\alpha_{n\wt\perp}\to\bs\alpha_{n\perp}$ as $\bh n_n\to\bh u_n$.

\section{Summary and discussion}\label{Conclusion}

In this paper, we have developed a manifestly Lorentz-covariant framework for analyzing electromagnetic fields generated by moving point charges. We systematically construct all relevant quantities in terms of covariant variables, including velocity, acceleration, and light-cone structure. This leads to a consistent and self-contained formulation of classical electromagnetism within the framework of special relativity.

A key contribution of this work is the derivation of the electric and magnetic fields in arbitrary inertial frames from the past light cone of each observer. The electromagnetic field at a given spacetime point is determined by the motion of source charges at the intersection of their worldlines with the past light cone of that point, ensuring causal consistency and Lorentz covariance. Our formulation avoids reliance on any particular inertial frame or coordinate choice, yielding expressions that make the observer-dependence of the electromagnetic field explicit. This clarifies the structure of relativistic field transformations and the interplay between geometry and dynamics in electromagnetic interactions.

We have also examined several physically relevant limits, including the non-relativistic regime, the ultra-relativistic case, and the field behavior along and transverse to the direction of motion. These analyses confirm the consistency of our formulation with familiar results—such as the recovery of Coulomb's law—and reveal distinct relativistic features such as field compression, directional enhancement, and asymmetry, which are absent in conventional treatments. This highlights the utility of our covariant formalism in bridging intuition across inertial frames.

Our formulation not only deepens the theoretical understanding of relativistic electromagnetism but also provides a solid foundation for future developments. These include analytical applications, numerical simulations, and educational tools aimed at visualizing field dynamics in a fully relativistic and covariant setting. We hope that this work will contribute both to a clearer theoretical perspective and to the development of computational and pedagogical methods for relativistic field theory. An explicit real-time implementation of this formalism for interactive visualization is presented in Ref.~\cite{nakayama2025seeing}.

\subsection*{Acknowledgement}

The work of K.O. is in part supported by JSPS KAKENHI Grant No.~23K20855.

\appendix
\section*{Appendix}

\section{General solution to relativistic Maxwell's equation}\label{general solution section}
In this section, we present a covariant derivation of the general solution to relativistic Maxwell's equation, based on the Green's function method in Lorenz gauge.

\subsection{Rewriting Maxwell's equation in Lorenz gauge}
\label{sec:Rewriting Maxwell's equation in Lorenz gauge}
In terms of the vector potential, Eq.~\eqref{matter equation of motion} becomes
\al{
\p^\mu\pn{\lv\p\cdot\lv A\fn{\lx}} - \Box A^\mu\fn{\lx}
	&=	{j^\mu\fn{\lx}\ov\epsilon_0c^2},
}
where $\lv\p\cdot\lv A\fn{\lx} := \p_\nu A^\nu\fn{\lx}$, and $\Box := \p^\nu\p_\nu = -\p_0^2 + \nab^2$ denotes the d'Alembertian.\footnote{
Here, it is understood that $\lv\p = \pn{\p^0,\p^1,\p^2,\p^3} = \pn{-\p_0,\p_1,\p_2,\p_3}$.
More explicitly, $\lv\p\cdot\lv A = \p_\mu A^\mu = {1\ov c^2}{\p\phi\ov\p t} + \nab\cdot\bs A$.
}

The field strength~\eqref{field strength defined} and the equation of motion~\eqref{matter equation of motion} are invariant
under the gauge transformation
\al{
A_\mu\fn{\lx}
	&\to	A'_\mu\fn{\lx}=A_\mu\fn{\lx}+\p_\mu\chi\fn{\lx}
}
with an arbitrary real scalar function $\chi\fn{\lx}$.
This gauge freedom allows us to impose the Lorenz gauge condition:
\al{
\lv\p\cdot\lv A\fn{\lx}
	&=	0,
}
which results in
\al{
-\Box A^\mu\fn{\lx}
	&=	{j^\mu\fn{\lx}\ov\epsilon_0c^2}.
}

\subsection{Green's function method for solving vector potential}
\label{sec:Green's function method for solving vector potential}
We review the derivation of the general solution~\eqref{general solution of A} to Eq.~\eqref{Maxwell's equation in Lorenz gauge}.

Using Green's function that satisfies
\al{
-\Box G\fn{\lx}
	&=	\delta^4\fn{\lx},
}
the general solution to Eq.~\eqref{Maxwell's equation in Lorenz gauge} can be written as
\al{
A^\mu\fn{\lx}
	&=	\int\df^4\green{\lv x'} G\fn{\lx - \green{\lv x'}}{j^\mu\fn{\green{\lv x'}}\ov \epsilon_0 c^2},
		\label{vector potential by Green's function}
}
where $\delta^4\fn{\lx}=\delta\fn{\xz}\delta^3\fn{\bx}=\delta\fn{\xz}\delta\fn{\orange{x^1}}\delta\fn{\orange{x^2}}\delta\fn{\orange{x^3}}$ is the spacetime Dirac delta function (distribution).

Now we outline the standard derivation of Green's function.
By the Fourier transform
\al{
G\fn{\lx}
	&=	\int{\df^4k\ov\pn{2\pi}^4}e^{i\lv k\cdot\lx}\wt G\fn{\lv k},
	\quad\pn{=	\int{\df k^0\ov2\pi}e^{-ik^0\xz}\int{\df^3\bs k\ov\pn{2\pi}^3}e^{i\bs k\cdot\bx}\wt G\fn{k^0,\bs k}}\\
\delta^4\fn{\lx}
	&=	\int{\df^4k\ov\pn{2\pi}^4}e^{i\lv k\cdot\lx},
}
we obtain
\al{
\wt G\fn{\lv k}
	=	{1\ov{\lv k}^2}={1\ov-\pn{k^0}^2+\bs k^2}.
}
Physically, $k^0$ corresponds to the angular frequency $\omega=ck^0$ of the electromagnetic field, namely the light.
Putting this back into the original expansion, we obtain
\al{
G\fn{\lx}
	&=	-{1\ov8\pi^2\ab{\bx}}\int_0^\infty\df k
		\pn{e^{ik\ab{\bx}}-e^{-ik\ab{\bx}}}
		\int{\df k^0\ov2\pi i}e^{-ik^0\xz}
		\pn{{1\ov k^0-k}-{1\ov k^0+k}}.
}
For the integration over $k^0$, on physical ground, we take the retarded Green's function that takes into account only the propagation of the light from the past to the future:\footnote{
When $\xz>0$, the complex $k^0$ integral is closed by the contour in the lower half plane, which picks up both the positive- and negative-energy poles at $k^0=\pm k-i\epsilon$.
When $\xz<0$, it is closed by that in the upper half plane, which picks up no pole and the integral becomes zero.
}
\al{
G_\tx{ret}\fn{\lx}
	&=	-{1\ov8\pi^2\ab{\bx}}\int_0^\infty\df k
		\pn{e^{ik\ab{\bx}}-e^{-ik\ab{\bx}}}
		\ub{\int{\df k^0\ov2\pi i}e^{-ik^0\xz}
			\pn{{1\ov k^0-k+i\epsilon}-{1\ov k^0+k+i\epsilon}}}_{
			\theta\pn{\xz}\pn{-e^{-ik\xz}+e^{ik\xz}}
			}\nn
	&=	{1\ov4\pi\ab{\bx}}\theta\fn{\xz}\sqbr{
			\delta\fn{\xz-\ab{\bx}}-\delta\fn{\xz+\ab{\bx}}
			}
	=	{1\ov4\pi\ab{\bx}}\delta\fn{\xz-\ab{\bx}},
	\label{retarded green's function}
}
where $\epsilon$ is a positive infinitesimal
and, in the second step, we used the Fourier integral representation of the delta function
$
\delta\fn{x}=\int_{-\infty}^\infty{\df k\ov2\pi}e^{ikx}$.

Putting the retarded Green's function~\eqref{retarded green's function} into Eq.~\eqref{vector potential by Green's function}, we get the general form of the vector potential:
\al{
	A^\mu\fn{\lx}
	&= {1\ov 4\pi \epsilon_0 c^2}\int\df^4\green{\lv x'} 
		\delta\fn{\xz-\green{x^{\pr0}} -  \ab{\bx - \green{\bs{x^\pr}}}}
	{j^\mu\fn{\green{\lv x'}}\ov\ab{\bx - \green{\bs{x^\pr}}}}.
		\label{A concrete in Appendix}
}
For brevity, we will sometimes write the integral as
\al{
\int_{\PLC\fn{\lx}} \df^3\green{\bs{x'}}\sqbr{\cdots}
	\,:=\,
		\int\df^4\green{\lv x'} 
		\delta\fn{\xz-\green{x^{\pr0}} -  \ab{\bx - \green{\bs{x^\pr}}}}
		\sqbr{\cdots}
}
such that
\al{
A^\mu\fn{\lx}
	&=	{1\ov 4\pi\epsilon_0c^2}\int_{\PLC\fn{\lx}} \df^3\green{\bs{x'}} {j^\mu\fn{\green{\lv x'}} \ov \left|\bx - \green{\bs x'}\right|}.
	\label{general solution of A in Appendix}
}

\section{Basic derivative formulae}\label{sec:Useful formulae}
For reader's ease, we list derivatives with respect to $\bx$,
\al{
{\p l_n^i\fn{\sn,\bx}\ov\p\oxj}
	&=	-\delta^{ij},\label{dlidxj}\\
{\p\ab{\bs l_n\fn{\sn,\bx}}\ov\p\oxi}
	&=	-\h l_n^i\fn{\sn,\bx},\\
{\p\h l_n^i\fn{\sn,\bx}\ov\p\oxj}
	&=	{-{\delta^{ij}}+\h l_n^i\fn{\sn,\bx}\h l_n^j\fn{\sn,\bx}\ov\ab{\bs l_n\fn{\sn,\bx}}},\\
{\p\gamma_n\fn{\sn,\bx}\ov\p\oxi}
	&=	{-u_n^i\fn{\sn}+\h l_n^i\fn{\sn,\bx}\pn{\bh l_n\fn{\sn,\bx}\cdot\bs u_n\fn{\sn}}\ov\ab{\bs l_n\fn{\sn,\bx}}}
}
and derivatives with respect to $\sn$,
\al{
{\p\bs l_n\fn{\sn,\bx}\ov\p\sn }
	&=	\bs u_n\fn{\sn },
		\label{dldtau}\\
{\p\ab{\bs l_n\fn{\sn,\bx}}\ov\p\sn }
	&=	\bh l_n\fn{\sn,\bx}\cdot\bs u_n\fn{\sn },
		\label{d abs l d tau}\\
{\p\bh l_n\fn{\sn,\bx}\ov\p\sn }
	&=	{\bs u_n\fn{\sn }-\pn{\bh l_n\fn{\sn,\bx}\cdot\bs u_n\fn{\sn }}\bh l_n\fn{\sn,\bx}\ov\ab{\bs l_n\fn{\sn,\bx}}},\\
{\p\gamma_n\fn{\sn,\bx}\ov\p\sn }
	&=	\alpha_n^0\fn{\sn }
		+\bh l_n\fn{\sn,\bx}\cdot\bs\alpha_n\fn{\sn }
		+{\bs u_n^2\fn{\sn }-\pn{\bh l_n\fn{\sn,\bx}\cdot\bs u_n\fn{\sn }}^2\ov\ab{\bs l_n\fn{\sn,\bx}}}.
		\label{d2r0dtau2}
}

\section{Spacetime differentiation on past light cone and of retarded vector potential}
\label{sec:Spacetime differentiation on past light cone and of retarded vector potential}
When we regard Eq.~\eqref{yet another trivial equation} as a relation among variables $\xz$, $\bx$, and $\sn$, we obtain
\al{
\df\xz-u_n^0\fn{\sn}\df\sn
	&=	\ub{\p\ab{\bs l_n\fn{\sn,\bx}}\ov\p\sn}_{\bh l_n\fn{\sn,\bx}\cdot\bs u_n\fn{\sn}}\df\sn+\sum_{i=1}^3\ub{\p\ab{\bs l_n\fn{\sn,\bx}}\ov\p\oxi}_{-\h l_n^i\fn{\sn,\bx}}\df\oxi,
}
which leads to
\al{
\gamma_n\fn{\sn,\bx}\df\sn
	&=	\df\xz+\sum_{i=1}^3\h l_n^i\fn{\sn,\bx}\df\oxi.
}
From this, we obtain
\al{
\left.{\p\snsx\ov\p\xz}\right|_{\bx\tx{ fixed}}
	&=	{1\ov\gamma_n\fn{\snsx,\bx}},&
\left.{\p\snsx\ov\p\oxi}\right|_{\xz,\,\orange{x^{j\black{(\neq i)}}}\tx{ fixed}}
	&=	{\h l_n^i\fn{\snsx,\bx}\ov\gamma_n\fn{\snsx,\bx}}.
	\label{dtaudx}
}

\subsection{Verification of derivatives on past light cone}
In the master formula~\eqref{final true master equation for vector potential}, the following equality holds by the definition of $\snsx$ as the solution to Eq.~\eqref{yet another trivial equation}:
\al{
\ab{\bs l_n\fn{\snsx,\bx}}=\xz-x_n^0\Fn{\snsx}.
	\label{on-PLC condition}
}
Using Eqs.~\eqref{dlidxj}--\eqref{d2r0dtau2} and \eqref{dtaudx}, we now verify that the spacetime derivatives of both sides of Eq.~\eqref{on-PLC condition} consistently coincide.

First, the derivative with respect to $\xz$ is given by
\al{
\left.{\p\ab{\bs l_n\fn{\snsx,\bx}}\ov\p\xz}\right|_{\bx\tx{ fixed}}
	&=	{\p\snsx\ov\p\xz}\left.{\p\ab{\bs l_n\fn{\sn,\bx}}\ov\p\sn}\right|_{\sn=\snsx}
	=	\left.{\bh l_n\fn{\sn,\bx}\cdot\bs u_n\fn{\sn }\ov\gamma_n\fn{\sn,\bx}}
		\right|_{\sn=\snsx},
		\label{time derivative of absolute ln}\\
\restrict{\p\Bigl(\xz-x_n^0\Fn{\snsx}\Bigr)\ov\p\xz}{\bx\tx{ fixed}}
	&=	1-{\p\snsx\ov\p\xz}\restrict{\df x_n^0\fn{\sn}\ov\df\sn}{\sn=\snsx}
	=	\left.\pn{1-{u_n^0\fn{\sn}\ov\gamma_n\fn{\sn,\bx}}}\right|_{\sn=\snsx}.
		\label{time derivative of time difference}
}
These expressions are seen to agree due to Eq.~\eqref{gamma_n defined}.

Next, the derivative with respect to $\oxi$ is computed as
\al{
\left.{\p\ab{\bs l_n\fn{\snsx,\bx}}\ov\p\oxi}\right|_{\xz,\,\orange{x^{j\black{(\neq i)}}}\tx{ fixed}}
	&=	{\p\snsx\ov\p\oxi}\restrict{\p\ab{\bs l_n\fn{\sn,\bx}}\ov\p\sn}{\sn=\snsx}
		+\restrict{\p\ab{\bs l_n\fn{\sn,\bx}}\ov\p\oxi}{\sn=\snsx}\nn
	&=	\left.\pn{
			\pn{\bh l_n\fn{\sn,\bx}\cdot\bs u_n\fn{\sn }}
			{\h l_n^i\fn{\sn,\bx}\ov\gamma_n\fn{\sn,\bx}}
			-\h l_n^i\fn{\sn,\bx}
			}\right|_{\sn=\snsx},
		\label{spatial derivative of absolute ln}\\
\left.{\p\Bigl(\xz-x_n^0\Fn{\snsx}\Bigr)\ov\p\oxi}\right|_{\xz,\,\orange{x^{j\black{(\neq i)}}}\tx{ fixed}}
	&=	-{\p\snsx\ov\p\oxi}\restrict{\df x_n^0\fn{\sn}\ov\df\sn}{\sn=\snsx}
	=	-\left.{u_n^0\fn{\sn}\h l_n^i\fn{\sn,\bx}\ov\gamma_n\fn{\sn,\bx}}\right|_{\sn=\snsx}.
}
Again, the results are found to be consistent.

We have thus confirmed that both the temporal and spatial derivatives of Eq.~\eqref{final true master equation for vector potential} are consistently handled under the identity~\eqref{on-PLC condition}.
\subsection{Spacetime derivatives of modified gamma factor}
For later use, we list the partial derivatives of the modified gamma factor restricted onto $\PLC\fn{\lx}$:
\al{
\restrict{\p\gamma_n\fn{\snsx,\bx}\ov\p\xz}{\bx\tx{ fixed}}
	&=	{\p\snsx\ov\p\xz}\restrict{\p\gamma_n\fn{\sn,\bx}\ov\p\sn}{\sn=\snsx}\nn
	&=	
		{1\ov\gamma_n\fn{\sn,\bx}}
		\Bigg[
			\alpha_n^0\fn{\sn }
			+\bh l_n\fn{\sn,\bx}\cdot\bs\alpha_n\fn{\sn }\nn
	&\phantom{={1\ov\gamma_n\fn{\sn,\bx}}\Bigg[}
			+{\bs u_n^2\fn{\sn }-\pn{\bh l_n\fn{\sn,\bx}\cdot\bs u_n\fn{\sn }}^2\ov\ab{\bs l_n\fn{\sn,\bx}}}
			\Bigg]
		\Bigg|_{\sn=\snsx},
		\label{x0 derivative of gamma}\\
\restrict{\p\gamma_n\fn{\snsx,\bx}\ov\p\oxi}{\xz,\,\orange{x^{j\black{(\neq i)}}}\tx{ fixed}}
	&=	{\p\snsx\ov\p \oxi}\restrict{\p\gamma_n\fn{\sn,\bx}\ov\p\sn}{\sn=\snsx}
		+\restrict{\p\gamma_n\fn{\sn,\bx}\ov\p \oxi}{\sn=\snsx}\nn
	&=	\Bigg\{
		{\h l_n^i\fn{\sn,\bx}\ov\gamma_n\fn{\sn,\bx}}
		\Bigg[
			\alpha_n^0\fn{\sn }
			+\bh l_n\fn{\sn,\bx}\cdot\bs\alpha_n\fn{\sn }\nn
	&\phantom{=\Bigg\{{\h l_n^i\fn{\sn,\bx}\ov\gamma_n\fn{\sn,\bx}}\Bigg[}
			+{\bs u_n^2\fn{\sn }-\pn{\bh l_n\fn{\sn,\bx}\cdot\bs u_n\fn{\sn }}^2\ov\ab{\bs l_n\fn{\sn,\bx}}}\Bigg]
			\nn
	&\phantom{=\Bigg\{}
		+{-u_n^i\fn{\sn}+\h l_n^i\fn{\sn,\bx}\pn{\bh l_n\fn{\sn,\bx}\cdot\bs u_n\fn{\sn}}\ov\ab{\bs l_n\fn{\sn,\bx}}}
		\Bigg\}
		\Bigg|_{\sn=\snsx}.
		\label{xi derivative of gamma}
}

\subsection{Spacetime derivatives of vector potential}

Using Eqs.~\eqref{dtaudx}, \eqref{time derivative of absolute ln}, and \eqref{x0 derivative of gamma}, we compute the time derivative of the vector potential~\eqref{final true master equation for vector potential}:
\al{
\p_0A^\mu\fn{\lx}
	&=	\sum_n{q_n\ov4\pi\epsilon_0c}
		\Bigg[
		{1\ov\gamma_n\Fn{\snsx,\bx}\ab{\bs l_n\Fn{\snsx,\bx}}}
		{\p u_n^\mu\Fn{\snsx}\ov\p \xz}\nn
	&\phantom{=\sum_n{q_n\ov4\pi\epsilon_0c}\Bigg[}
		-{u_n^\mu\fn{\snsx}\ov\gamma_n^2\Fn{\snsx,\bx}\ab{\bs l_n\Fn{\snsx,\bx}}}{\p\gamma_n\Fn{\snsx,\bx}\ov\p \xz}\nn
	&\phantom{=\sum_n{q_n\ov4\pi\epsilon_0c}\Bigg[}
		-{u_n^\mu\fn{\snsx}\ov\gamma_n\Fn{\snsx,\bx}\ab{\bs l_n\Fn{\snsx,\bx}}^2}{\p\ab{\bs l_n\Fn{\snsx,\bx}}\ov\p \xz}
		\Bigg]\nn
	&=	\sum_n{q_n\ov4\pi\epsilon_0c}\Bigg[
			{\alpha_n^\mu\fn{\sn }\ov\gamma_n^2\fn{\sn,\bx}\ab{\bs l_n\fn{\sn,\bx}}}\nn
	&\phantom{=	\sum_n{q_n\ov4\pi\epsilon_0c}\Bigg[}
				-{u_n^\mu\fn{\sn }\ov\gamma_n^3\fn{\sn,\bx}\ab{\bs l_n\fn{\sn,\bx}}}
					\Bigg(
						\alpha_n^0\fn{\sn }
		+\bh l_n\fn{\sn,\bx}\cdot\bs\alpha_n\fn{\sn }\nn
	&\phantom{=	\sum_n{q_n\ov4\pi\epsilon_0c}\Bigg[-{u_n^\mu\fn{\sn }\ov\gamma_n^3\fn{\sn,\bx}\ab{\bs l_n\fn{\sn,\bx}}}\Bigg(}
		+{\bs u_n\fn{\sn}\cdot\pn{\bs u_n\fn{\sn}+u_n^0\fn{\sn}\bh l_n\fn{\sn,\bx}}\ov\ab{\bs l_n\fn{\sn,\bx}}}
				\Bigg)
				\Bigg]
		\Bigg|_{\sn=\snsx},
}
where we used $\bs u_n^2-\pn{\bh l_n\cdot\bs u_n}^2+\gamma_n\pn{\bh l_n\cdot\bs u_n}=\bs u_n\cdot\pn{\bs u_n+u_n^0\bh l_n}$ due to Eqs.~\eqref{normalization of u} and \eqref{gamma_n defined}, in the last step.

Similarly, using Eqs.~\eqref{dtaudx}, \eqref{spatial derivative of absolute ln}, and \eqref{xi derivative of gamma}, the spatial derivative reads
\al{
\p_iA^\mu\fn{\lx}
	&=	\sum_n{q_n\ov4\pi\epsilon_0c}
		\Bigg[
		{1\ov\gamma_n\Fn{\snsx,\bx}\ab{\bs l_n\Fn{\snsx,\bx}}}
		{\p u_n^\mu\fn{\snsx}\ov\p \oxi}\nn
	&\phantom{=\sum_n{q_n\ov4\pi\epsilon_0c}\Bigg[}
		-{u_n^\mu\fn{\snsx}\ov\gamma_n^2\Fn{\snsx,\bx}\ab{\bs l_n\Fn{\snsx,\bx}}}{\p\gamma_n\Fn{\snsx,\bx}\ov\p \oxi}\nn
	&\phantom{=\sum_n{q_n\ov4\pi\epsilon_0c}\Bigg[}
		-{u_n^\mu\fn{\snsx}\ov\gamma_n\Fn{\snsx,\bx}\ab{\bs l_n\Fn{\snsx,\bx}}^2}{\p\ab{\bs l_n\Fn{\snsx,\bx}}\ov\p \oxi}
		\Bigg]\nn
	&=	\sum_n{q_n\ov4\pi\epsilon_0c}\Bigg[
				{\h l^i\fn{\sn,\bx}\alpha_n^\mu\fn{\sn}\ov\gamma_n^2\fn{\sn,\bx}\ab{\bs l_n\fn{\sn,\bx}}}
				+{u_n^i\fn{\sn}u_n^\mu\fn{\sn}\ov\gamma_n^2\fn{\sn,\bx}\ab{\bs l_n\fn{\sn,\bx}}^2}\nn
	&\phantom{=	\sum_n{q_n\ov4\pi\epsilon_0c}\Bigg[}
				-{\h l_n^i\fn{\sn,\bx}u_n^\mu\fn{\sn}\ov\gamma_n^3\fn{\sn,\bx}\ab{\bs l_n\fn{\sn,\bx}}}
					\pn{
						\alpha_n^0\fn{\sn}
						+\bh l_n\fn{\sn,\bx}\cdot\bs\alpha_n\fn{\sn}
						-{1\ov\ab{\bs l_n\fn{\sn,\bx}}}
						}
				\Bigg]\Bigg|_{\sn=\snsx},
}
where we used $\bs u_n^2
		-\pn{\bh l_n\cdot\bs u_n}^2
		+\gamma_n\pn{\bh l_n\cdot\bs u_n}
		-\gamma_nu_n^0=-1$ due to Eqs.~\eqref{normalization of u} and \eqref{gamma_n defined}, in the last step.

\bibliographystyle{JHEP}
\bibliography{refs}

\end{document}